\documentclass[11pt]{article}
\usepackage{moriond,epsfig}

\usepackage[T1]{fontenc}
\usepackage[latin9]{inputenc}
\usepackage{xspace}
\usepackage{amsfonts}
\usepackage{amsmath}
\usepackage{amssymb}
\usepackage{epsfig}
\usepackage{graphicx}
\usepackage{psfrag}
\usepackage{bm}

\def\be{\begin{equation}}
\def\ee{\end{equation}}
\def\bea{\begin{eqnarray}}
\def\eea{\end{eqnarray}}

\def\keV{\mathrel{\,{\rm keV}}} 
 
\def\GeV{\mathrel{\,{\rm GeV}}}

\def\lsim{\mathrel{\lower4pt\hbox{$\sim$}} 
\hskip-9.5pt\raise1.6pt\hbox{$<$}\;} 
 
\def\gsim{\mathrel{\lower4pt\hbox{$\sim$}} 
\hskip-12.5pt\raise1.6pt\hbox{$>$}\;} 
 
\begin{document}
\vspace*{4cm}
\title{  Dark Matter from the Inert Doublet Model }

\author{  Laura Lopez Honorez}

\address{Service de Physique Th\'eorique, Universit\'e Libre de Bruxelles,\\
 CP225, Bld du Triomphe, 1050 Brussels, Belgium}

\maketitle\abstracts{
The  Inert Doublet Model  is an extension of the Standard Model including
one extra ``Inert scalar doublet'' and an exact $Z_2$ symmetry.  
The ``Inert scalar'' provides a new  candidate
 for dark matter.
 We present a systematic analysis of the dark matter abundance assuming the
 standard freeze-out mechanism and 
investigate the potentialities for direct and gamma indirect detection. 
We show that the dark matter candidate saturates the WMAP dark matter density   in two rather
separate mass ranges, one between 40 and 80 GeV, the other one over  400 GeV.
We also show that the model 
should be within the range of future experiments, 
like GLAST and EDELWEISS II or  ZEPLIN. 
}
 
\section{Introduction} 

Many evidences for the existence in the universe of dark  matter has been put forward
over the years.
It can be inferred from  the dynamics of galaxies and of clusters of galaxies, from analysis
of the CMBR, from structure formation, etc. 
The question that arises then is what is the nature of dark matter and which extension of the Standard Model do
we have to consider in order to account for these observations?
A profusion of models  of dark particles have been 
proposed over the years and it is 
much hoped that present and forthcoming experiments will 
throw some light 
on the matter (for a review, see for instance \cite{Bertone:2004pz}).

In these proceedings, we study a simple extension of the
Standard Model with one extra scalar doublet and an exact $Z_2$
symmetry.
In this framework, the candidate for dark matter is one of the two neutral scalars
arising from the extra doublet.
The latter was called ``Inert doublet'' by  Barbieri
{\it et al} in \cite{Barbieri:2006dq} because it has no  direct coupling to matter fields.
However it couples to  the standard  gauge   fields. 
The phenomenology of its neutral
and charged components is quite simple and yet very rich. 

This is not the only attractive feature of the model.
As was pointed out in \cite{Barbieri:2006dq}, the Inert Doublet  Model
(IDM) could allow for a Higgs mass up to 500 GeV still fulfilling the
LEP Electroweak Precision Test measurements. 
Here, we will only consider a Higgs with a mass of 
120 GeV. 
In the reference paper \cite{LopezHonorez:2006gr}, one can find  a more detailed study of the IDM with Higgs masses of
120 GeV and 200 GeV
(See also \cite{Gustafsson:2007pc}  where the authors considered Higgs masses
up to 500 GeV).
Another interesting aspect of models like the IDM  is that it could pave the
way toward an understanding of the relation between the abundance of dark and
ordinary matter (see {\it e.g.} \cite{Gu:2007mi}).

\section{Short description of the Model}

 The IDM is a particular two Higgs doublet model, in which one of the doublet, $H_1$,
plays the role of the standard Brout-Englert-Higgs doublet while the second
one, $H_2$, is  the source for dark matter candidates.
In order to guarantee the stability of the dark matter particles,  one invoke
a $Z_2$ symmetry  under
which all Standard Model fields are even and
$$ 
H_1 \rightarrow H_1 \;\; \mbox{\rm and}\;\; 
H_2\rightarrow - H_2.
$$ 
This discrete 
symmetry also prevents the appearance 
of flavor changing neutral currents in this model.
Moreover, we assume that $Z_2$ is not
 spontaneously broken.
 This model was first introduced by Deshpande and Ma \cite{PhysRevD.18.2574}
 (see also \cite{Ma:2006km}),
 and the dark matter aspect was recently discussed by Cirelli {\it et al} \cite{Cirelli:2005uq} and Barbieri
{\it et al} \cite{Barbieri:2006dq}.
 Their initial purpose and some of
their assumptions were nevertheless not exactly identical.
In addition, the neutral scalar
reaching the dark matter WMAP abundance was found to be in the  mass range of
60 to 75 GeV for Barbieri
{\it et al} \cite{Barbieri:2006dq} while for  Cirelli {\it et al}
\cite{Cirelli:2005uq} it was of order of 430 GeV.
We first study the details of the  model before to elucidate this apparent incompatibility.

  The most general potential of the model 
can be written as 
\begin{equation} 
\label{potential} 
V = \mu_1^2 \vert H_1\vert^2 + \mu_2^2 \vert H_2\vert^2  + \lambda_1 \vert H_1\vert^4 + 
 \lambda_2 \vert H_2\vert^4 + \lambda_3 \vert H_1\vert^2 \vert H_2 \vert^2 
 + \lambda_4 \vert H_1^\dagger H_2\vert^2 + {\lambda_5\over 2} \left[(H_1^\dagger H_2)^2 + h.c.\right]
\end{equation} 
The vacuum expectation value of $H_1$ is given by
$\langle H_1\rangle = {v\over \sqrt 2}$
 with $ v =\sqrt{ -\mu_1^2/\lambda_1}= 248$ GeV, while assuming for simplicity
 $\mu_2^2 > 0$,
  we have
$ 
\langle H_2\rangle = 0. 
$ 
The mass of the Higgs 
particle $h$  is 
$M_h^2 = - 2 \mu_1^2 \equiv 2 \lambda_1 v^2$ 
while the mass of the 
charged, $H^+$, and two neutral, $H_0$ and $A_0$, components of the field
$H_2$ are given by 
\begin{eqnarray} 
M_{H^+}^2 &=& \mu_2^2 + \lambda_3 v^2/2\cr 
M_{H_0}^2 &=& \mu_2^2 +  (\lambda_3 + \lambda_4 + \lambda_5) v^2/2\cr 
M_{A_0}^2 &=& \mu_2^2 +  (\lambda_3 + \lambda_4 - \lambda_5) v^2/2. 
\label{masses} 
\end{eqnarray} 

For appropriate quartic couplings, $H_0$ or $A_0$ is the 
lightest component of the $H_2$ doublet.
 In the absence of any other lighter $Z_2$-odd field, 
either one is a candidate for dark matter. 
For definiteness we choose $H_0$. All our 
conclusions are unchanged 
if the dark matter candidate is $A_0$ instead. 
Following \cite{Barbieri:2006dq}, we parameterize the contribution from 
symmetry breaking to the mass of $H_0$ by $\lambda_L=(\lambda_3 + \lambda_4 + 
\lambda_5)/2$, which is also the coupling constant between the Higgs field $h$
  and our dark matter candidate $H_0$. 

\section{Dark matter abundance}

 \begin{figure} 
\begin{center} 
\begin{tabular}{cc} 
\includegraphics[width=0.47\textwidth]{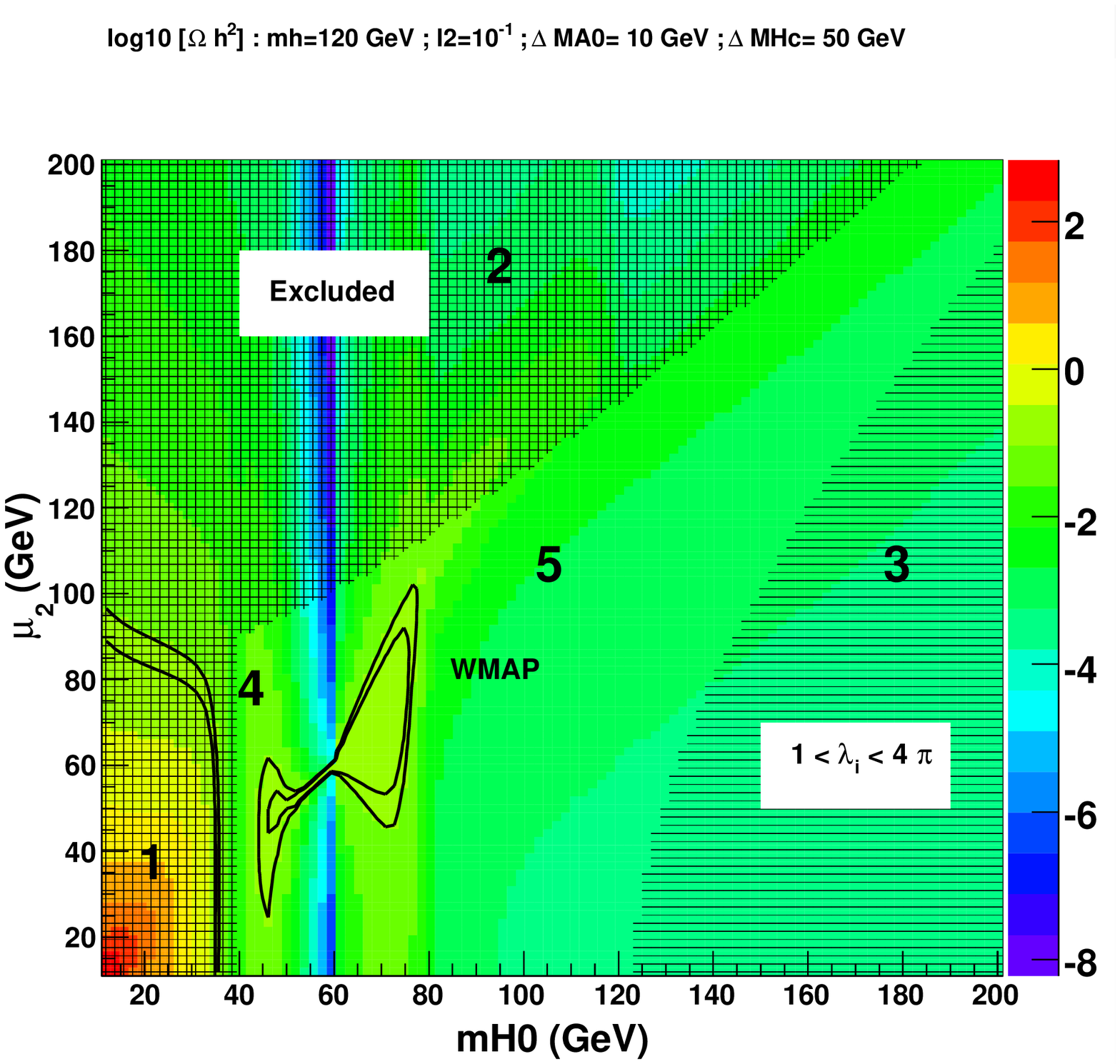}& 
\includegraphics[width=0.47\textwidth]{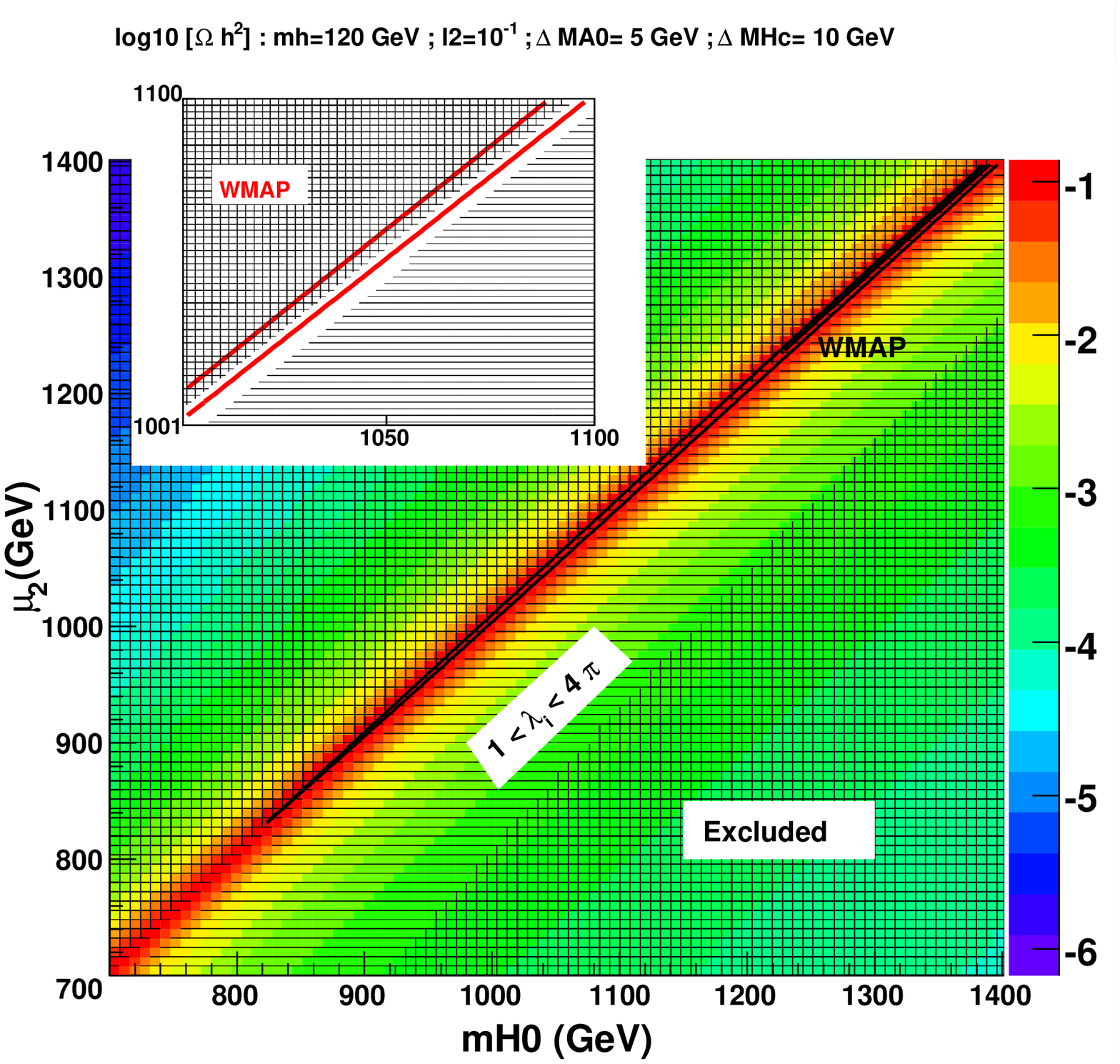}
\end{tabular} 
\caption{ Relic density contours in the $(M_{H_0},\mu_2)$ plane for
  $M_h=120$. Left: low mass regime with Inert scalars mass differences
  $\Delta MA_0=10 \GeV$ and $\Delta MH^+=50 \GeV$  .  Right: high mass regime with Inert scalars mass differences
  $\Delta MA_0=5 \GeV$ and $\Delta MH^+=10 \GeV$.
} 
\label{fig:abundance}
\end{center}
 \end{figure} 

As in \cite{Barbieri:2006dq} and in
\cite{Cirelli:2005uq},  we  consider a thermal production of the  cold relic
$H_0$.  
We have computed the relic 
abundance of $H_0$ using micrOMEGAs2.0, a new and versatile 
package for the numerical calculation 
of dark matter abundance from thermal freeze-out \cite{Belanger:2006is}. 

 We first present  the results for fixed Inert scalars mass differences in the
contour plots of figure~\ref{fig:abundance}.
We work in the ($M_{H_0},\mu_2$) plane, as a result
the diagonal line corresponds to $\lambda_L =0$, {\em i.e. to no coupling} between $H_0$ and the Higgs boson. 
Away from this line, $\lambda_L$ increases, with $\lambda_L <0$ (resp. $\lambda_L > 0$) above (resp. below) 
the diagonal. 
Also, we write $\Delta MA_0= M_{A_0}-M_{H_0}$ and  $\Delta MH_c=
M_{H^+}-M_{H_0}$.

The shaded 
areas in the plots  correspond to
regions that are excluded by several constraints.
In order not to conflict with  LEP data, the mass of the $H^+$ should be
larger than 79.3 GeV and $M_{H^0} + M_{A^0}\lsim M_Z$.
These constraints translate into the excluded region 1 on the plot on the
left of figure~\ref{fig:abundance}.
 The vacuum stability constraint contributes largely to the exclusion of $\lambda_L<0$ couplings. 
This corresponds to 
shaded area in the domains $\mu_2 > M_{H_0}$ in the plots of
figure~\ref{fig:abundance}.
The remaining shaded regions are excluded due to large couplings  
 $|\lambda_i|~>~4\pi$. Moreover regions where the couplings range as
$1~<~|\lambda_i|~<~4\pi$,  which is still tolerable,  are shown with horizontal lines.
The areas between two dark lines correspond to regions 
of the parameter space such that $0.094<\Omega_{DM}h^2< 0.129$, the range of 
 dark matter energy densities consistent with WMAP data.

We immediately see that there are two qualitatively distinct regimes, depending on whether the $H_0$ 
is lighter than the $W$ and $Z$ and/or the Higgs boson.
For the low mass regime, let us study figure~\ref{fig:abundance}, the plot on the left.
    The two processes  relevant below the $W$, $Z$ or $h$ threshold are the $H_0$ annihilation 
through the Higgs and $H_0$ coannihilation with $A_0$ through $Z$ exchange\footnote{$H_0\,H^+$ coannihilation is suppressed for our choice of $\Delta MH_c$.}. 
Both give fermion-antifermion pairs, the former predominantly into $b\bar b$. 
Coannihilation into a $Z$ may occur provided $\Delta MA_0$ is not
too important, 
roughly $\Delta MA_0$ must be of order of $T_{fo}\sim M_{H_0}/25$. 
As the mass of $H_0$ goes above $W$, $Z$ or $h$ threshold, $H_0$ annihilation
into  $WW$, $ZZ$ and $hh$ become 
increasingly efficient, an effect which strongly suppresses the $H_0$ relic density.
The region 4 of figure~\ref{fig:abundance} , corresponding to  $M_{H_0} \in [40,80]$ GeV, appears to be the only region consistent with WMAP data.

For the high mass regime, we can derive the general trends from
figure~\ref{fig:abundance}, the plot on the right.
No new annihilation channel opens if $M_{H_0}$ is heavier than 
the Higgs or the gauge bosons. There are then essentially two kinds  of 
processes which control both the abundance: the annihilation into two gauge bosons, dominant if 
$\mu_2<M_{H_0}$, and the annihilation into two 
 Higgs, which dominates if $\mu_2>M_{H_0}$. 
 Coannihilation plays little role.

The abundance of dark matter is suppressed over most of the area of the
plot because of large quartic coupling effects on the cross-sections. 
Let us emphasize that in this regime, it is possible to reach agreement with WMAP data, but only at the price of
some fine tuning.
We need to keep  the mass splittings between the components of
$H_2$ relatively small. 
First because large mass splittings correspond to large couplings and second because 
the different contributions to the annihilation cross-section must be suppressed at the same location, 
around  $\lambda_L=0$ ({\it i.e.} $M_{H_0}\simeq \mu \simeq M_{A_0}\simeq
M_{H_+}$ in this case).
As it can be seen  in figure~\ref{fig:abundance}, the plot on the right, the area consistent with WMAP corresponds to the narrow region around
the diagonal  with
$M_{H_0}\gsim 800$ GeV for $\Delta MA_0 = 5$ GeV and  $\Delta MH_c = 10$ GeV.
Notice that this behavior is limited by  the unitarity bound on the total annihilation
cross-section \cite{Griest:1989wd} which constrains the mass of the dark
particle to  be $M_{H_0}\lsim 120$ TeV. 

\begin{figure}
\begin{center} 
\begin{tabular}{cc} 
\includegraphics[width=0.47\textwidth]{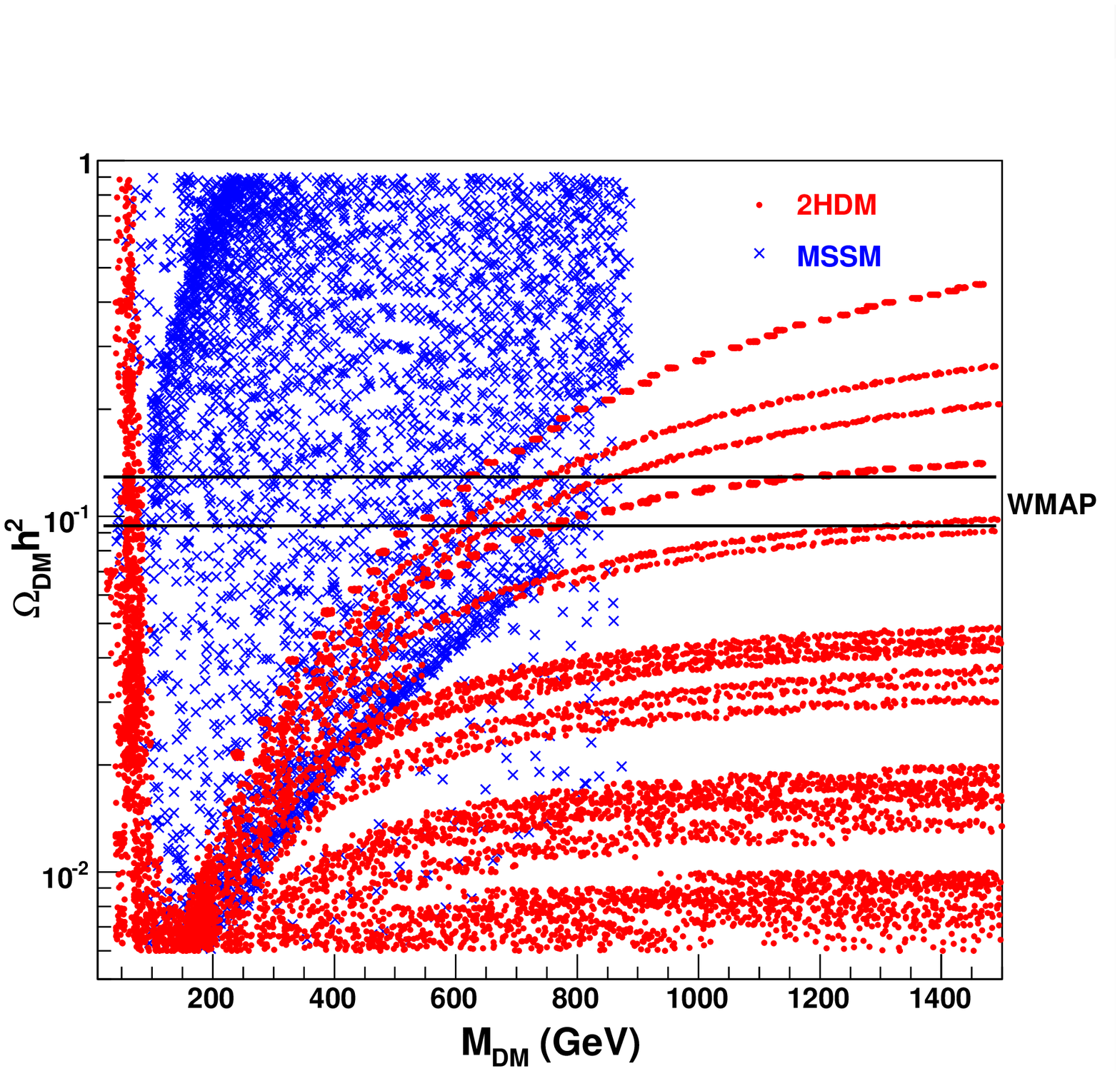}
&\includegraphics[width=0.47\textwidth]{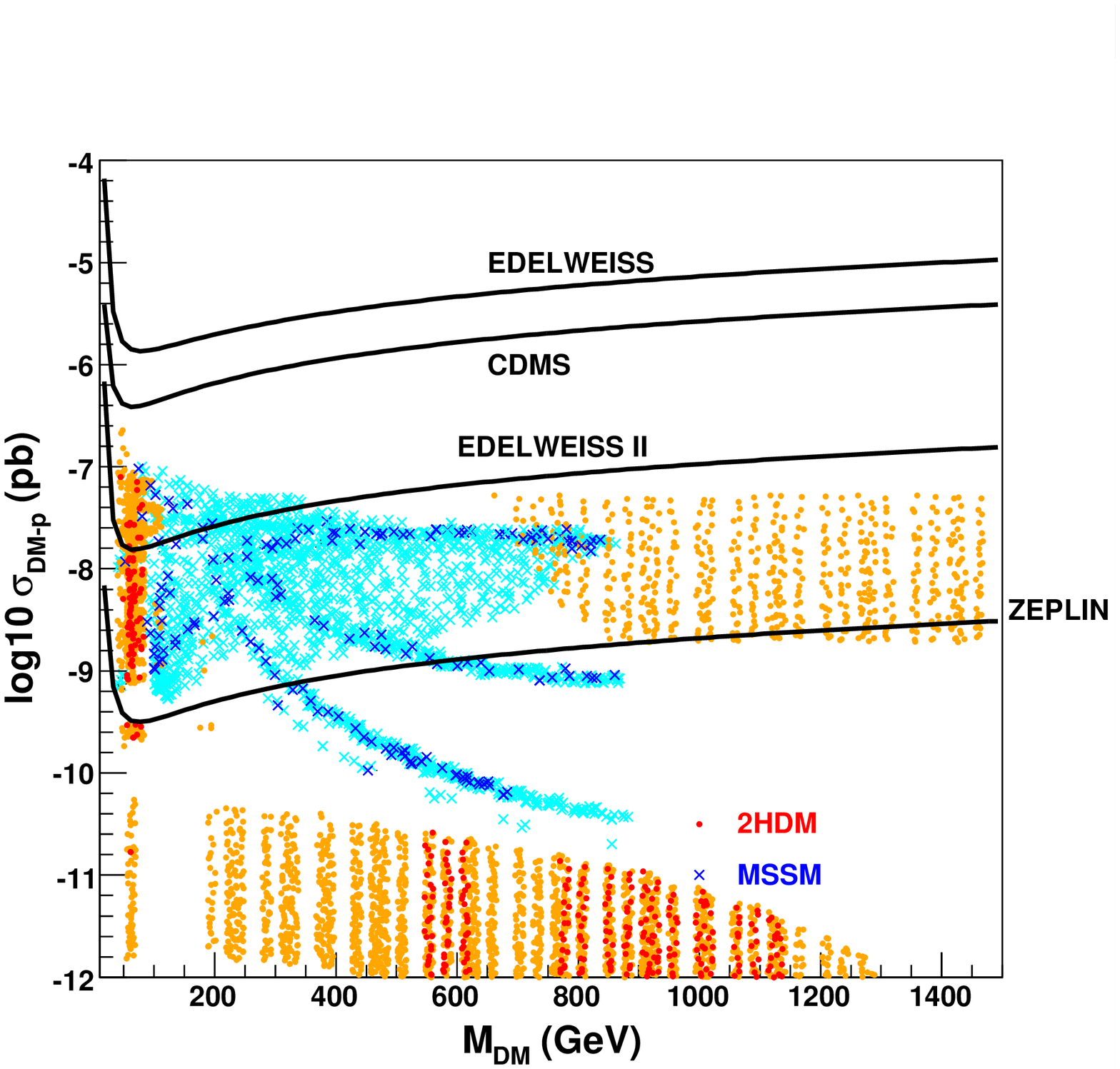}
 \end{tabular} 
\caption{Left:  Relic density and  Right: scattering cross-section intervening
  in direct detection searches,  all as a function
  of the mass of dark matter and comparison with the MSSM. 
 For the direct detection plot, the light colors correspond to $0.01<\Omega_{DM}h^2<0.3$, while the dark
  colors correspond to $0.094<\Omega_{DM}h^2<0.129$.
} 
\label{fig:scan}
\end{center} 
\end{figure} 

In figure \ref{fig:scan}, the plot on the left, we show a scatter plot of $\Omega_{DM} h^2$  as a function of the mass of the dark matter candidate $M_{DM}$ for a
 fair sample of IDMs (scanning on several Inert scalar mass
 splittings) and, for the sake of comparison, for the MSSM.
 We clearly see the two regimes (low mass and high mass) of the IDM that may give
 rise to a relevant relic density
({\it i.e} near WMAP).  The MSSM models have 
a more continuous behavior, with 
${\cal O}(100$ GeV) dark matter masses. 
As a conclusion, the IDM provides dark matter candidates with masses as
 small as 40 GeV and as large as 600 GeV in contrast with the MSSM more
 concentrated around $\sim$ 100 GeV.

\section{ Direct  detection}

Direct detection searches look for signals of dark matter in low background detectors
trying to measure the energy deposited by the scattering of a dark
 matter particle with a nucleus of the detector.   
Assuming that the main interaction contributing to the  $H_0$-quarks
interaction is the spin independent $H_0 q \xrightarrow{h}H_0 q$ interaction\footnote{The experiments have reached such a level of sensitivity
  that the $Z$ exchange contribution $H_0 q \xrightarrow{Z}A_0 q$ is excluded by the current 
experimental limits 
 \cite{Barbieri:2006dq}. Consequently, to forbid $Z$ exchange by kinematics, the mass of the $A_0$ particle must be
larger
than the mass of $H_0$ by a few $100 {\keV}$.
} it can be shown \cite{Barbieri:2006dq} that the $H_0$ elastic scattering cross
section off a proton scales like 
\begin{equation}
\sigma_{H_0-p}\propto {\lambda_L}^2/({M_{H_0}M^2_h})^2.
\label{eq:DD}
 \end{equation}

In figure \ref{fig:scan}, the  plot on the right, we show a scatter plot 
 of $\log_{10}\sigma_{DM-p}$ as a function of $M_{DM}$ for the IDMs considered in the abundance plot of figure \ref{fig:scan} that account for $0.01<\Omega_{DM}h^2<0.3$. 
We see that the low mass regime candidates could be
detected by future experiments such as EDELWEISS II  or by the  ton sized experiments such as ZEPLIN.
For the higher mass regime however, there is no hope for future detection in
low background detector.
Indeed, the WMAP requirement for dark matter relic density constrains the
$\lambda_L$ couplings to be vanishing while the same couplings drive the
amplitude of the matter-$H_0$ scattering cross-section.

\section{ Indirect  detection}

\begin{figure}
\begin{center} 
\begin{tabular}{cc} 
\includegraphics[width=0.47\textwidth]{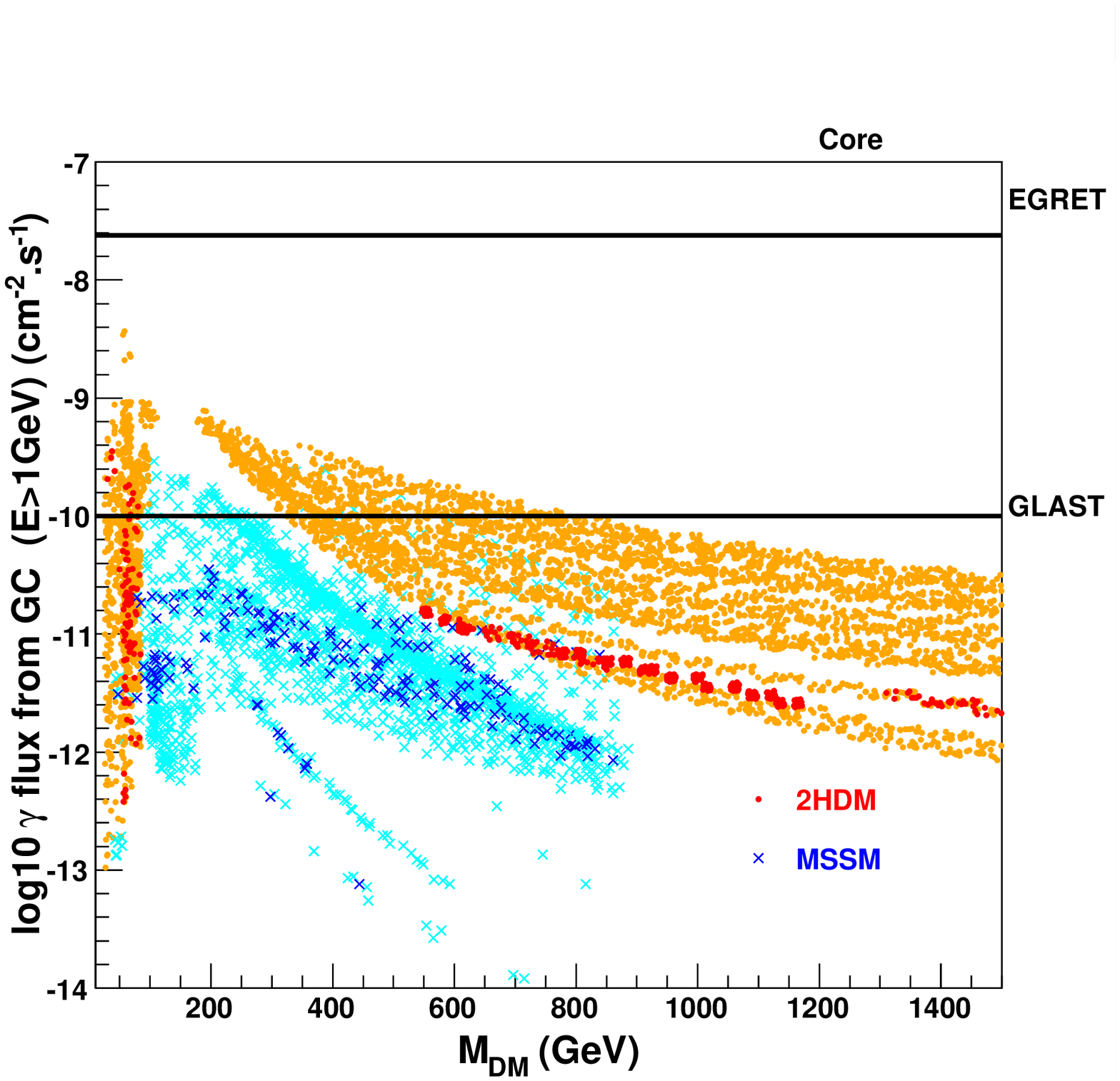}
&\includegraphics[width=0.47\textwidth]{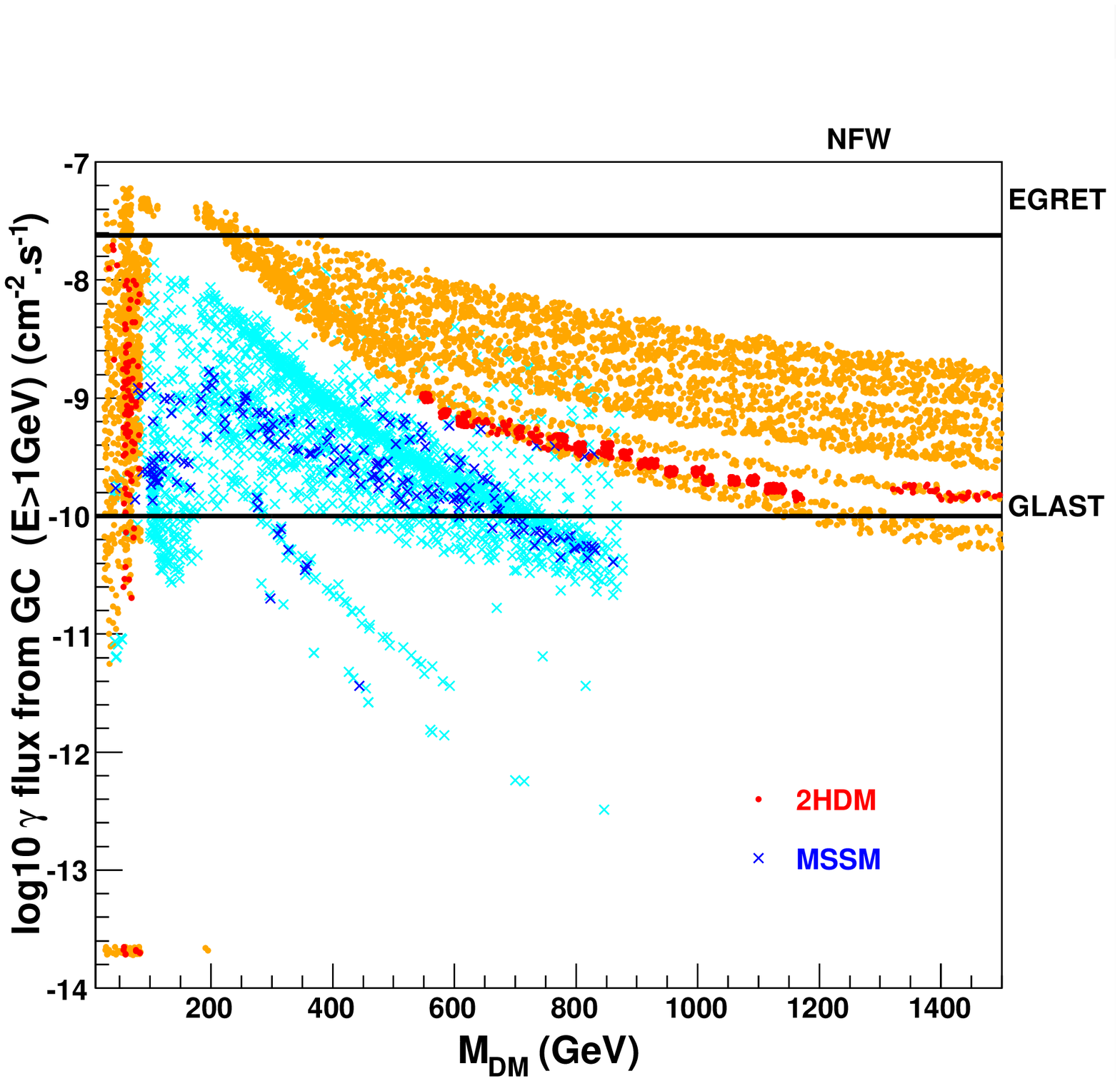}
 \end{tabular} 
\caption{Integrated gamma-ray flux from the Galactic Center (GC) resulting
from dark matter annihilation as a function of the mass of the dark matter
candidate for the same sample of models than for direct detection in
Fig.\ref{fig:scan}.
Again, the light colors correspond to $0.01<\Omega_{DM}h^2<0.3$, while the dark
  colors correspond to $0.094<\Omega_{DM}h^2<0.129$.
For the plot on the left, we the took an isothermal profile (flat profile), for the plot on
the right, we took a NFW profile (more steeper at Galactic Center). 
} 
\label{fig:IDscan}
\end{center} 
\end{figure} 

The measurement of secondary particles coming from dark matter 
annihilation in the halo of the Galaxy is another promising way of deciphering the
nature of dark matter. 
Let us emphasize that  this 
possibility depends however not only on the properties of the dark matter particle, through its annihilation cross-sections, 
but also on the astrophysical 
assumptions made concerning the distribution of dark matter in the halo that
supposedly surrounds our Galaxy.
 
In Figure \ref{fig:IDscan}, we show the $\log$ of the    produced
gamma-ray flux from dark matter annihilation at the
 Galactic Center $\Phi_\gamma$ as a function of $M_{DM}$ for the same
 sample of models than for direct detection.
We computed the gamma-ray flux for the plot on the left assuming an isothermal
dark matter density profile while for the plot on the right we assumed a
Navarro, Frank and White (NFW) profile.
The main difference between these two profiles is the slope of the dark matter
density as a function of the galactic radius in the central part of the Galaxy.
 The isothermal profile is flat while the NFW profile is more cuspy ({\it i.e.}
steeper).
We see that for steeper profile the gamma-ray flux is larger.

The particle physics dependence of $\Phi_\gamma$   also clearly  show up  in
figure~\ref{fig:IDscan}.
Indeed we see that
$\Phi_\gamma$ behaves differently in the low and the high mass regime of the
IDM given that the processes contributing to the annihilation cross-section
are different.
 Moreover, notice that  the IDM dark matter candidates have typically higher detection rates than the neutralino in 
SUSY models, especially at high mass. 
Let us stress that the figures for indirect detection were obtained taking into
account annihilation processes at three level only (see
\cite{Gustafsson:2007pc}, for a recent study of the IDM including processes at
one-loop). 
It can be inferred from  figure~\ref{fig:IDscan} that 
 the IDM 
can give the right relic abundance in a range of parameters which will be
 probed by GLAST for NFW dark matter profiles.
GLAST will however have no chance to observe the gamma-ray flux produced by
 annihilating $H_0$ at the Galactic Center  for flatter profiles such as the
 isothermal one.

\section{Conclusion}

We carried out a rather detailed analysis of the IDM as a dark matter model  assuming
the 
standard freeze-out mechanism.
We recovered the results of Barbieri {\it et al.} and Cirelli {\it et al.}
which a priori did not seem to match.
This is because the IDM provides dark matter candidates in two rather
separate mass ranges, one between 40 and 80 GeV, the other one over  400 GeV.
The physics driving the existence of  dark matter in these regions of the parameter space is
quite different.

We have also investigated the prospects for direct and indirect detection searches.
Concerning direct detection searches, the low mass regime candidates should
be detected with the futures ton sized experiments while the high mass regime
will stay out of reach.
For indirect detection searches we looked at the gamma-ray flux generated at  the
Galactic Center by dark matter annihilation.
Whatever the dark matter density profile
assumed, we have come to the conclusion that the Inert
scalars have typically   higher detection rates than the neutralino in 
SUSY models, especially at high mass.
Moreover, the IDM could be probed by the future GLAST experiment.

\section*{Acknowledgments}

This work is supported by the FNRS, the I.I.S.N. and the
Belgian Federal Science Policy (IAP 5/27).
\section*{References}
\bibliographystyle{unsrt}
 \bibliography{biblio} 
\end{document}